\documentclass[a4paper]{aa}  
\usepackage{graphicx}
\usepackage{longtable}
\usepackage{txfonts}

\usepackage{natbib}
\begin{document}

\title{Stellar coronal magnetic fields and star-planet interaction}

    \titlerunning{Stellar magnetic fields and exoplanets}
    \authorrunning{A. F. Lanza}

%   \subtitle{}

   \author{A.~F.~Lanza}

   \offprints{A.~F.~Lanza}

   \institute{INAF-Osservatorio Astrofisico di Catania, Via S. Sofia, 78 
               -- 95123 Catania, Italy \\ 
              \email{nuccio.lanza@oact.inaf.it}    
             }

   \date{Received ... ; accepted ... }

    \abstract{Evidence of magnetic interaction between late-type stars and {{close-in}} giant planets is  provided by the observations of stellar hot spots rotating synchronously with the planets and showing an enhancement of {{ chromospheric and X-ray fluxes}}. Possible photospheric signatures of such an interaction have also been reported.}{We investigate star-planet interaction in the framework of a magnetic field model of a stellar corona, considering the interaction between the coronal field and that of a planetary magnetosphere moving through the corona. {{  This is motivated, among others,  by the difficulty of accounting for the  energy budgets of the interaction phenomena with previous models.}} }{A linear force-free model is applied to describe the coronal field and study the evolution of its total magnetic energy and relative helicity according to the boundary conditions at the stellar surface and the effects related to the planetary motion through the corona.}{The energy budget of the star-planet interaction is discussed assuming that the planet may trigger a release of the  energy of the coronal field by decreasing its relative helicity. The observed intermittent character of the star-planet  interaction is explained by a topological change of the stellar coronal field, induced by a variation of its relative helicity.  The model predicts the formation of many prominence-like structures in the case of highly active stars owing to the accumulation of matter evaporated from the planet inside an azimuthal flux rope in the outer corona. Moreover, the model can explain why stars accompanied by close-in planets have a higher X-ray luminosity than those with distant planets. It predicts that the best conditions to detect radio emission from the exoplanets and their host stars are achieved when the field topology is characterized by field lines connected to the surface of the star, leading to a chromospheric hot spot rotating synchronously with the planet.}{The main predictions of the model can be verified with present observational techniques, by a simultaneous monitoring of the chromospheric flux and X-ray (or radio) emission, and spectropolarimetric observations of the photospheric magnetic fields.  }

   \maketitle

%________________________________________________________________

\section{Introduction}

More than 350 extrasolar giant planets are presently 
known\footnote{See a web catalogue at: http://exoplanet.eu/}, among which $\sim 25$ percent have a
projected orbital semimajor axis smaller than 0.1 AU. Such planets are expected 
to interact  with their host stars, not only through tides, but  also
with other mechanisms, possibly associated with  magnetic fields. Specifically, reconnection between the coronal field of the host star and the magnetic field of the planet is expected to release heat,  produce hydromagnetic waves, and accelerated particles that may be conveyed onto the stellar chromosphere producing a localized enhanced emission \citep{Cuntzetal00,Ipetal04,Preusseetal06,McIvoretal06,CranmerSaar07}.  Indeed, \citet{Shkolniketal05,Shkolniketal08} show that HD~179949 and \object{$\upsilon $~Andromedae}  have chromospheric hot spots  that rotate with the orbital periods of their inner planets.  Two other stars, 
\object{HD 189733} and \object{$\tau$ Bootis}, show  evidence of  an excess of chromospheric variability, probably due to flaring, that is modulated with the orbital periods of their respective planets \citep[cf. also ][]{Walkeretal08}.  
Modelling such features can allow us to obtain information on the magnetic fields of the planets, although in an indirect way \citep[cf. ][]{Shkolniketal08}, which is of great importance to constrain their internal structure and evolution as well as to characterize a possible habitability  in the case of rocky planets. A recent step forward in the understanding of the star-planet magnetic interaction (hereinafter SPMI) is the work of \citet{Lanza08}. { It applies force-free and non-force-free coronal field models  to account for the observed phase lags between planets and synchronous chromospheric hot spots. Moreover, it presents conjectures about a possible influence of a close-in planet on the hydromagnetic dynamo action occurring in its host star.
In the present paper, we shall extend that work discussing   topology, total energy, and relative magnetic helicity of the coronal field  and investigating their role in the effects associated with the motion of a hot Jupiter inside a stellar corona.} 
We shall propose a new mechanism to account for the energy budget of SPMI and its intermittent nature, as revealed by the latest observations (see Sect.~\ref{observations}). 
Moreover, we shall discuss how the different coronal field topologies and the presence of a 
planet embedded in the corona  may affect the formation of prominence-like structures as well as the X-ray and radio emissions of a star. 

\section{Observations}
\label{observations}

Evidence of SPMI was reported by { \citet{Shkolniketal03,Shkolniketal05,Shkolniketal08}} who observed  chromospheric hot spots rotating with the orbital period of the  planets in \object{HD~179949} and \object{$\upsilon$~And} instead of their respective rotation periods. The spots were not located at the subplanetary point, i.e., along the line joining the centre of a star with its planet, but lead the planet by $\sim 70^{\circ}$ in the case of 
\object{HD~179949} and $\sim 170^{\circ}$ in the case of \object{$\upsilon$ And}, respectively. 
The flux irradiated by the hot spot was greater in the case of \object{HD~179949}, with an excess power of $\sim 10^{20}$ W, corresponding to that of a large solar flare. The hot spot synchronized with the planet was observed in four out of six seasons in HD~179949, suggesting that it is not a steady phenomenon, but { may have a lifetime not exceeding $300-400$ days}. A similarly intermittent interaction is suggested by the observations of $\upsilon$~And, leading \citet{Shkolniketal08} to propose that these on/off SPMI { transitions} are related to a variation of the configuration of the coronal magnetic field of the host stars, possibly connected to their activity cycles. For the planet hosts \object{HD~189733} and \object{HD~73256}, \citet{Shkolniketal08} suggest a correlation of the amplitude of the intranight variability of the Ca~II~K line core flux with the orbital phase of the planet, with a maximum of activity leading the planet by $\approx 70^{\circ}$ in the case of \object{HD~189733}. This may be due to some chromospheric flaring activity synchronized with the planet.  

The X-ray flux coming from \object{HD~179949} appears to be modulated with the orbital period of the planet with a variation  of $\approx$ 30 percent \citep{Saaretal07}.
A statistical analysis of a  sample of late-type stars  reveals that those hosting a hot Jupiter closer than 0.15~AU  have, on the average, an X-ray flux $ \approx 3-4$ times greater than stars hosting planets farther than $1.5$ AU \citep{Kashyapetal08}. This suggests that a close-in giant planet may increase the X-ray activity level of its host star. 

Circumstantial evidence of  planet-induced photospheric magnetic activity in  stars hosting hot Jupiters has been reported in \object{$\tau$ Bootis} by \citet{Walkeretal08} and in CoRoT-2a and CoRoT-4a by \citet{Lanzaetal09a}, \citet{Paganoetal09}, and \citet{Lanzaetal09b}, respectively. For \object{$\tau$ Boo} there is evidence of an active region leading the planet by about $70^{\circ}$ whose signatures have been observed both in the Ca~II~K line core flux and in the wide-band optical flux as monitored by the MOST (Microvariability and Oscillation of STars) satellite. 
The region shows a rotational modulation of $\sim 1$ mmag. In 2004 
it resembled a dark spot of variable depth, while in 2005 it varied between bright and dark.
{Since $\tau$ Boo has an average rotation period of 3.3 days, synchronized with the orbital period of the planet, and displays a surface differential rotation comparable to  that of the Sun \citep{Catalaetal07}, it is the constancy of the phase shift between the spot and the planet  that gives support to a possible SPMI in this case. As a matter of fact, the hot spot can be traced back to 2001, thanks to  a previous Ca~II~K line flux monitoring}. 

The planet host \object{CoRoT-4a} as a spectral type similar to $\tau $ Boo and its rotation appears to be  synchronized on the average with that of its planet. Modelling two months of uninterrupted CoRoT observations, \citet{Lanzaetal09b} found evidence of a persistent active region at the subplanetary longitude. Five months of observations of \object{CoRoT-2a} revealed a modulation of the total spotted area with a cycle close to ten synodic periods of the hot Jupiter with respect to the stellar rotation period \citep{Lanzaetal09a}, again suggesting some kind of SPMI. 
{ \citet{Paganoetal09} found that the variance of the stellar flux was modulated in phase with the planetary orbit, with  minimum variance at phase $0.3-0.4$ and maximum at phase $0.8-0.9$.}
Finally, \citet{Henryetal02} found evidence of a photospheric spot in \object{HD~192263} that in two observing seasons rotated with the orbital period of its planet  for at least three rotation cycles \citep[see ][]{Santosetal03}. 

This preliminary evidence of photospheric { cool} spots synchronized with a hot Jupiter is impossible to explain in the framework of a magnetic reconnection model { because reconnection leads to energy release and thus  heating of the atmosphere. Conversely, it suggests that the planet may affect in some way the stellar dynamo action or the emergence of magnetic flux, as conjectured by \citet{Lanza08}}. 

{ Spectropolarimetric observations can be applied to map photospheric fields outside starspots and indeed for $\tau$ Boo a sequence of maps has been obtained that suggests that the magnetic activity cycle of the star is as short as $\sim 2$ years, instead of $22$ years as in the Sun
\citep{Donatietal08,Faresetal09}.} 

Solar system giant planets emit in the radio domain, mainly via the electron-cyclotron maser mechanism. Most of the power is coherently radiated near the cyclotron frequency $f_{\rm ce} = 2.8 B$ MHz, where $B$ is the intensity of the planet's magnetic field in Gauss. In the case of Jupiter, the maximum of the flux falls around 40 MHz. Solar system planets follow a scaling law that relates their emitted radio power to the power supplied  by the impinging solar wind
at their magnetospheric boundary. 
 Generalizing  that scaling law, it is possible to predict exoplanetary radio emission powers  
{ \citep[cf., e.g.,][]{Stevens05,Zarka07,JardineCameron08}}. For $\tau$ Boo and some other systems, fluxes  between 30 and 300 mJy, i.e., within the detection limits of some of the largest radio telescopes, have been predicted for favourable conditions. Nevertheless, no positive detection has been reported yet, not only in the case of $\tau$ Boo, observed at several epochs at 74 MHz with upper limits between 135 and 300 mJy \citep{LazioFarrell07}, but also for \object{$\epsilon$ Eridani} and \object{HD~128311} at 150 MHz \citep{GeorgeStevens07} for which tight upper limits of $10-20$ mJy were derived, and for \object{HD~189733}, for which an upper limit of $\sim 80$ mJy was reached over the 307-347 MHz range \citep{Smithetal09}. It is hoped that the next generation of low-frequency radio telescopes can lower those limits by at least one order of magnitude allowing us to clarify whether the missed detections are due to a lack of sensitivity or  a  lack of emission from the exoplanets at those frequencies. As a matter of fact, if hot Jupiters have magnetic field strengths comparable to that of Jupiter, { i.e., $B \sim 14.5$ G at the poles (with 
$B \sim 4.3$~G at the planet's equator)}, most of their radio emission falls at frequencies lower than those of the above mentioned surveys. Another possibility is that the emission is beamed out of the line of sight, or that the stellar wind power (kinetic and/or magnetic) is significantly smaller than assumed.

\section{Coronal field model}
% \subsection{Modelling chromospheric hot spots}
\label{model}

We adopt  a spherical polar coordinate frame having its  origin at the baricentre of the host star and the polar axis along the stellar rotation axis. The radial distance from the origin is indicated with $r$,  the colatitude measured from the North pole with $\theta$, and the azimuthal angle with $\phi$. The planet orbit is assumed circular with a semimajor axis $a$ and lying in the equatorial plane of the star\footnote{{ About 70 percent of the hot Jupiters within 0.1~AU from their host stars have a measured orbital eccentricity lower than 0.05, consistent with a circular orbit (see http://exoplanet.eu/).}}. 

Close to a star, the magnetic pressure in the corona is much greater than the plasma pressure and the gravitational force, so we can assume that the corona is in a force-free magnetohydrostatic balance, i.e., the current density ${\vec J}$ is everywhere parallel to the magnetic field ${\vec B}$, viz. ${\vec J} \times {\vec B} = 0$. This means that $\nabla \times {\vec B} = \alpha {\vec B}$, with { the force-free parameter} $\alpha$ constant along each field line \citep{Priest82}. If $\alpha$ is uniform in the stellar corona, the field is called a linear force-free field and it satisfies the vector Helmoltz equation $\nabla^{2} {\vec B} + \alpha^{2} {\vec B} = 0$. Its solutions in spherical geometry have been studied by, e.g.,  \citet{Chandrasekhar56} and \citet{ChandrasekharKendall57}. 

Linear force-free fields are particularly attractive in view of their mathematical symplicity and their minimum-energy properties in a finite domain, as shown by \citet{Woltjer58}. Specifically, in ideal magnetohydrodynamics, the minimum energy state of a magnetic field in a finite domain is a linear force-free state set according to the boundary conditions and the constrain posed by the conservation of the magnetic helicity. In a confined stellar corona, magnetic dissipation is localized within thin current sheets whose global effect is that of driving the field configuration toward the minimum energy state compatible with the conservation of the total helicity, the so-called Taylor's state, which is a linear force-free field \citep[see, e.g., ][ for details]{HeyvaertsPriest84,Berger85}. Only in  large flares the total helicity is not conserved due to strong turbulent dissipation and ejection of magnetized plasma. 

The total helicity of a confined magnetic structure constrains  its free energy, that is the energy that can be released in a magnetic dissipation process. It is the difference between the initial energy of the field, which is usually in a non-linear force-free state (i.e., with a non-uniform $\alpha$), and the energy of the linear force-free field satisfying the same boundary conditions and having the same total helicity \citep[see, e.g., ][]{RegnierPriest07}. Therefore, a decrease of the total helicity produced by a change of the boundary conditions or a strong magnetic dissipation will in general make available more free energy for the heating of the corona. 
 
To model the magnetic interaction between the stellar coronal field and a close-in planet, we consider only the dipole-like component (i.e., with a radial order $n=1$) of the linear force-free solution of \citet{ChandrasekharKendall57} because it has the slowest decay with distance from the star and therefore leads to the strongest interaction. Moreover, since the observations of star-planet interaction show a hot spot rotating with the orbital period of the planet and do not show the periodicity of stellar rotation, an axisymmetric field (i.e., with an azimuthal degree $m=0$) is a good approximation to describe the interaction, as discussed by \citet{Lanza08}. 

Our linear force-free field can be expressed in the formulism of \citet{Flyeretal04} as: 
\begin{equation}
{\vec B} = \frac{1}{r \sin \theta} \left[ \frac{1}{r} \frac{\partial A}{\partial \theta} \hat{\vec r} - \frac{\partial A}{\partial r} \hat{\vec \theta} + \alpha A \hat{\vec \phi} \right],
\label{field_express}
\end{equation}
where $A(r, \theta)$ is the flux function of the field. 
Magnetic field lines lie  over surfaces of constant $A(r, \theta)$, 
as can be deduced by noting that ${\vec B} \cdot \nabla A = 0$. The flux function for our dipole-like field geometry is  $A(r, \theta) = B_{0} R^{2} g(q) \sin^{2} \theta$, where
$2 B_{0}$ is the magnetic field intensity at the North pole of the star, $R$ the star's radius and the function $g(q)$ is defined by:
\begin{equation}
g (q) \equiv \frac{[b_{0} J_{-3/2}(q) + c_{0} J_{3/2}(q)]\sqrt{q}}{[b_{0} J_{-3/2}(q_{0}) + c_{0} J_{3/2}(q_{0})] \sqrt{q_{0}}},
\end{equation}
where $b_{0}$ and $c_{0}$ are free constants, $J_{-3/2}$ and $J_{3/2}$ are Bessel functions of the first kind of order $-3/2$ and $3/2$, respectively, $q \equiv |\alpha | r$, and $q_{0} \equiv |\alpha | R$. Making use of Eq.~(\ref{field_express}), the magnetic field components are:
\begin{eqnarray}
B_{r} & = & 2B_{0} \frac{R^{2}}{r^{2}} g(q) \cos \theta, \nonumber \\
\label{field_conf}
B_{\theta}  & = & -B_{0} |\alpha | \frac{R^{2}}{r} g^{\prime} (q)  \sin \theta, \\
B_{\phi} & = & \alpha B_{0} \frac{R^{2}}{r} g(q)  \sin \theta \nonumber,  
\end{eqnarray}
where $g^{\prime} (q) \equiv dg/dq$. 
A linear force-free field as given by Eqs.~(\ref{field_conf}) extends to infinity with an infinity energy. We consider its restriction to the radial domain $q_{0} \leq q \leq q_{\rm L}$, where $q_{\rm L}$ is the first zero of $g(q)$, in order to model the inner part of the stellar corona where magnetic field lines are closed, as discussed in \citet{Lanza08} \citep[see ][ for the boundary conditions at $r=r_{\rm L} \equiv q_{\rm L}/ |\alpha | $]{Chandrasekhar56}. 

The magnetic field geometry specified by Eqs.~(\ref{field_conf}) depends on two independent parameters, i.e., $\alpha$ and $b_{0}/c_{0}$. They can be derived from the boundary conditions at the stellar photosphere, i.e., knowing the magnetic field ${\vec B}^{(s)}(\theta, \phi) $ on the surface at $r=R$. Using the orthogonality properties of the basic poloidal and toroidal fields \citep[see ][]{Chandrasekhar61}, we find: 
\begin{eqnarray}
\frac{ 8 \pi}{3} B_{0} R^{2} &  = & \int_{\Sigma(R)} B_{\rm r}^{(s)} \cos \theta d \Sigma,  \nonumber \\
\label{BC_eqs}
  \frac{ 8 \pi}{3} |\alpha | B_{0} R^{3} g^{\prime} (q_{0}) & = & - \int_{\Sigma(R)} B_{\theta}^{(s)} \sin \theta d \Sigma,  \\
\frac{ 8 \pi}{3} \alpha  B_{0} R^{3} & = & - \int_{\Sigma(R)} B_{\phi}^{(s)} \sin \theta d \Sigma, \nonumber 
\end{eqnarray}  
where  $\Sigma (R)$ is the spherical surface of radius $R$, and $d \Sigma = R^{2} \sin \theta d\theta d \phi$. 

The magnetic energy $E$ of the field confined between the spherical surfaces $r=R$ and $r=r_{\rm L}$ can be found from Eq.~(79) in \S~40 of \citet{Chandrasekhar61}:
\begin{equation}
E = E_{\rm p} \left\{ 2 + q_{0} q_{\rm L} [g^{\prime}(q_{\rm L})]^{2} - q_{0}^{2} [g^{\prime}(q_{0})]^{2} - q_{0}^{2} \right\}, 
\label{Benergy}
\end{equation}
where $E_{\rm p} \equiv \frac{4 \pi}{3 \mu} R^{3} B_{0}^{2}$ is the energy of the potential dipole field with the same radial component at the surface $r=R$, and $\mu$ is the magnetic permeability. The relative magnetic helicity $H_{\rm R}$, as defined by \citet{Berger85}, can be found from his Eq.~(19) and is:
\begin{equation}
H_{\rm R} = B_{0}^{2} R^{4} \left[ 2 g^{\prime}(q_{0}) + \frac{8 \pi}{3} \frac{E}{q_{0} E_{\rm p}} \right] \frac{|\alpha|}{\alpha}. 
\label{Bhelic}
\end{equation} 
Note that the field obtained by changing the sign of $\alpha$ has the same poloidal components $B_{\rm r}$ and $B_{\theta}$, and energy $E$, while the toroidal component $B_{\phi}$ and the relative helicity $H_{\rm R}$ become opposite. 

In the limit  $q \ll 1$, i.e., close to the stellar surface, neglecting terms
of the order $O(q^{2})$, we find that $g(q) \sim q_{0}/q$ and $g^{\prime}(q) \sim -q_{0}/q^{2}$, independently of $b_{0}$ and $c_{0}$. From Eqs.~(\ref{field_conf}), we see that 
$B_{\rm r}$ and $B_{\theta}$ close to the star are similar to the analogous components of the  potential dipole field with the same radial component at the surface.

It is also interesting to study the limit $\alpha \rightarrow 0$ of our force-free field. In this limit $r_{\rm L} \rightarrow \infty $ and the radial extension of the field grows without bound. Nevertheless, the field tends to become indistinguishable from the corresponding potential field  and $E \rightarrow E_{\rm p}$. In the same limit, the relative magnetic helicity grows without bound because $q_{0} \rightarrow 0$ in Eq.~(\ref{Bhelic}), but the helicity density (i.e., the magnetic helicity per unit volume) tends to zero \citep[see ][ for further description of this limit state]{Berger85,ZhangLow05}. 

For a finite $\alpha$, $E > E_{\rm p}$ because the potential field has the minimum energy for a given $B_{\rm r}^{(s)}$. If we consider all magnetic fields with one end of their field lines  anchored at $r=R$ and the other out to infinity, satisfying the same boundary conditions of our field at $r=R$, the field with the lowest possible energy is called the Aly field and its energy $E_{\rm Aly} = 1.66 E_{\rm p}$ \citep[see ][]{Flyeretal04}. We assume that the Aly energy is an upper bound for the energy of our field because it is the lowest energy allowing the field to open up all its lines of force out to infinity driving a plasma outflow 
similar to a solar coronal mass ejection.  

From a topological point of view, the fields obtained from Eqs.~(\ref{field_conf}) can be classified into two classes. If the function $g(q)$ decreases monotonously in the interval 
$q_{0} \leq q \leq q_{\rm L}$, all field lines  are anchored at both ends on the boundary $r=R$. On the other hand, if the function has a relative minimum (and a relative maximum) in that interval,  the field contains an azimuthal rope of flux located entirely in the $r > R$ space and running around the axis of symmetry. The magnetic field configurations considered by \citet{Lanza08} to model SPMI are of the first kind; an example of a field configuration containing an azimuthal flux rope will be discussed in Sect.~\ref{intermittency}. An analogous topological classification holds in the case of the non-linear force-free fields considered by \citet{Flyeretal04}. 

{ Note that the photospheric magnetic field components can be measured by means of spectropolarimetric techniques if the star rotates fast enough ($v \sin i \geq 10-15$~km~s$^{-1}$) as shown in the case of, e.g., $\tau$ Boo by \citet{Catalaetal07} and \citet{Donatietal08}. 
Therefore, Eqs.~(\ref{BC_eqs}) can be applied to derive the parameters of the coronal field model and its topology, as we shall show in Sect.~\ref{intermittency}}.

\section{Applications}
\label{application1}
In this Section we shall consider some applications of the above model for the stellar coronal field to the { phenomena} introduced in Sect.~\ref{observations}.

\subsection{Power dissipated in SPMI}

{ The power needed to explain the excess flux  from a  chromospheric hot spot or an X-ray emission synchronous with the planet is of the order of $10^{20}-10^{21}$ W. Magnetic reconnection between the stellar coronal field and the planetary field at the boundary of the planetary magnetosphere is not sufficient to account for such a power, as we show in Sect.~\ref{energy_budget1}. Therefore, we propose an interaction mechanism  that may be capable of sustaining that level of  power in Sect.~\ref{energy_budget2}}.

\subsubsection{A simple magnetic reconnection model}
\label{energy_budget1}

 The first mechanism proposed to account for the energy budget of SPMI is reconnection between the planetary and the stellar magnetic fields  at the boundary of the planetary magnetosphere. This boundary is characterized by a balance between the magnetic pressure of the coronal field and that of the planetary field.
The ram pressure is negligible because the planet is inside the region where the stellar wind speed is subalfvenic \citep[cf., e.g., ][]{Preusseetal05}, and the orbital velocity of the planet is about one order of magnitude smaller than the Alfven velocity \citep{Lanza08}. Therefore, assuming a planetary field with a dipole geometry, the radius of the planetary magnetosphere $R_{\rm m}$, measured from the centre of the planet, is given by:
\begin{equation}
R_{\rm m} = R_{\rm pl} \left[ \frac{B(a, \frac{\pi}{2})}{B_{\rm pl}} \right]^{-\frac{1}{3}}, 
\end{equation} 
where $R_{\rm pl}$ is the radius of the planet, $B(a, \frac{\pi}{2})$ is the { coronal field of the star on the  stellar equatorial plane} at  $r=a$, and $B_{\rm pl}$ is the magnetic field strength at the poles of the planet. 

\begin{figure}[t]
\includegraphics[width=8cm,height=8cm]{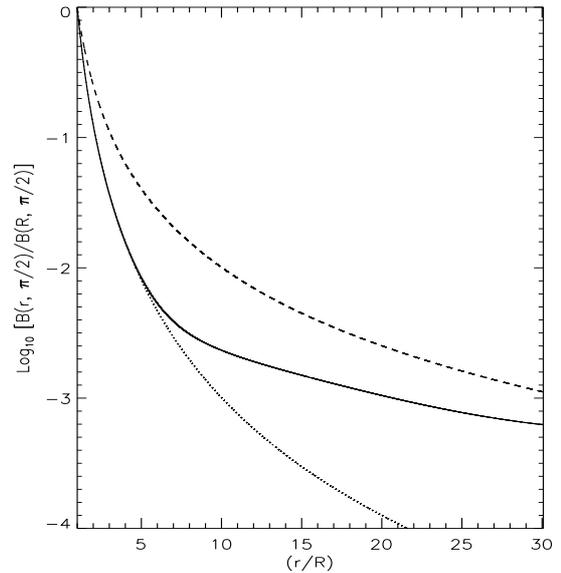}% {field_mod.ps} % {mag_hd179949_neu.ps}
\caption{The modulus of the magnetic field on the equatorial plane of the star normalized to its value at the surface vs. the radial distance from the star for $b_{0}/c_{0}=-1.1$ and $\alpha=-0.12 R^{-1}$ (solid line). For comparison, we plot also the case of a potential field (dotted line) and of a field decreasing as $r^{-2}$ (dashed line). }
\label{magneticf1}
\end{figure}

We specialize our considerations for the SPMI model of \object{HD~179949} developed by \citet{Lanza08} that assumes $b_{0}/c_{0}= -1.1$ and $\alpha=-0.12 R^{-1}$ to explain the phase lag between the chromospheric hot spot and the planet. The magnetic field intensity  vs. the distance from the star is plotted in Fig.~\ref{magneticf1}. { The field strength decreases as that of a potential dipole field close to the star, i.e., where $g(q) \sim q_{0}/q$ (cf. Sect.~\ref{model}), then  it decreases more slowly for $r > (6-7) R$ because the decrease of $g(q)$  becomes less steep (cf., e.g., Fig.~\ref{gplots}).}

At the distance of the planet, i.e., $a=7.72 R$, the field is reduced by a factor of $\sim 300$ with respect to its value at the stellar surface. Assuming a mean field at the surface $B_{0} = 10 $~G \citep[cf. ][]{Donatietal08},  the field at the boundary of the planetary magnetosphere is $B(a, \frac{\pi}{2}) \sim 0.03$~G.  Adopting a planetary field 
$B_{\rm pl}= 5$~G \citep[cf., e.g., ][]{Griessmeieretal04} and a radius of the planet equal to that of Jupiter, we find 
$R_{\rm m} \sim 5.5 R_{\rm pl} = 3.8 \times 10^{8}$~m. The power dissipated by magnetic reconnection can be estimated as:
\begin{equation}
P_{\rm d} \simeq \gamma \frac{\pi}{\mu} [B(a, \frac{\pi}{2})]^{2} R_{\rm m}^{2} v_{\rm rel}, 
\label{rec_power}
\end{equation} 
where $ 0 < \gamma < 1$ is a factor that depends on the angle between the interacting magnetic field lines \citep[see, e.g., ][]{Priest03}, and $v_{\rm rel}$ is the relative velocity between the planet and the stellar coronal field. Adopting the parameters for HD~179949 and $\gamma = 0.5$, we find $P_{\rm d} \simeq 1.3 \times 10^{17}$ W, which is insufficient by at least a factor of $10^{3}$. Since $P_{\rm d} \sim 
B^{4/3} B_{\rm pl}^{2/3}$, to explain the observed power the stellar surface field should be increased to
$B_{0} \sim 180 $ G, which is  too high in view of the mean values measured at the surface of 
\object{$\tau$ Boo} \citep{Donatietal08} that is a faster rotator than \object{HD~179949}, but with a comparable X-ray luminosity \citep{Kashyapetal08}. \citet{Christensenetal09} suggest that the surface fields of hot Jupiters may be up to $5-10$ times stronger than that of Jupiter. However, if we adopt
$B_{\rm pl} = 75$~G and consider the $ B_{\rm pl}^{2/3}$ dependence, we obtain a  power increase by only a factor of 6. 

\subsubsection{Magnetic energy release induced by a hot Jupiter}
\label{energy_budget2}

To solve the energy problem, it is important to note that the energy released at the reconnection site is only a  fraction of the energy that is actually available. The emergence of magnetic flux from the convection zone and the photospheric motions lead to a continuous accumulation of energy in the coronal field, independently of the presence of a planet. Our conjecture is that a close-in  planet triggers a release of such an accumulated energy, modulating the chromospheric and coronal heating with the orbital period of the planet. 

As a matter of fact, the  reconnection between the coronal and the planetary field lines has remarkable consequences for the topological structure of the coronal field. The final configuration of the coronal field after reconnection is in general a non-linear force-free one, i.e., $\alpha$ is no longer spatially uniform. This happens because the field  must simultaneously satisfy the boundary conditions at the photosphere, which impose the value of $\alpha$ close to the star, and those at the reconnection site close to the planet which in general will not be compatible with a uniform value of $\alpha $.  Since $\alpha$ must be constant along a given field line, the field re-arranges itself into a configuration with different values of $\alpha$ along different field lines to satisfy the boundary conditions.  We know from  Woltjer theorem that the energy of the field in such a state is greater than that of the linear force-free field with the same magnetic  helicity and  boundary conditions at the stellar surface. In other words, some energy can be released if the field makes a transition to this linear force-free state, restoring its unperturbed configuration. Such a transition can be stimulated by the fact that the helicity in the region where the field has reconnected is lower than the initial helicity because reconnection processes lead to a steady dissipation of helicity. Considering the volume $V(t)$ occupied at any given time $t$ by the plasma where reconnection has just occurred, the rate of helicity change inside $V(t)$ is given by Eq.~(13) of 
\citet{HeyvaertsPriest84}: 
\begin{equation}
\frac{D H_{\rm R}}{Dt} \! = \! \int_{S(V)}\! ({\vec A} \cdot {\vec v_{\rm rel}}) ({\vec B} \cdot {\vec d \vec S}) +\! \int_{S(V)} \! \frac{\vec A \times \vec J}{\sigma} \cdot \vec d \vec S - 2\! \int_{V} 
\! \frac{\vec B \cdot \vec J}{\sigma} d V, 
\end{equation}
where $S(V)$ is the surface bounding the volume $V(t)$, $\vec A$ is the vector potential of the magnetic field, viz. $ \nabla \times \vec A = \vec B$, and $\sigma$ is the electric conductivity of the plasma. We assume that the reconnected field is in a force-free state, thus $\mu \vec J = \alpha \vec B$, and choose a gauge transformation for the vector potential  to have $\vec A = \alpha^{-1} \vec B$. Since $ \vec v_{\rm rel} \cdot \vec B \sim 0$ in the reconnected region, we have: 
\begin{equation}
\frac{D H_{\rm R}}{Dt} \simeq -2 \langle \alpha^{-1} \rangle \mu \int_{V} \frac{J^{2}}{\sigma} dV, 
\label{helic_diss_rr}
\end{equation}
where $\langle \alpha^{-1} \rangle$ is some average value of $\alpha^{-1}$  over the volume $V(t)$. 
Eq.~(\ref{helic_diss_rr}) shows that the absolute value of the helicity decreases steadily inside the volume $V(t)$  following the motion of the reconnected region through the stellar corona. 

We conjecture that once a magnetic energy release is triggered by such a decrease of helicity in the reconnection volume, it extends to the whole flux tube connecting the planetary magnetosphere with the stellar surface and proceeds faster and faster thanks to a positive feed-back between energy release and helicity dissipation. In other words, the fast and localized energy release produces a turbulent plasma which in turn enhances turbulent dissipation of magnetic energy and helicity giving rise to a self-sustained process. 

An order-of-magnitude estimate of the helicity dissipation rate can be based on Eq.~(\ref{helic_diss_rr}), now considering turbulent dissipation; it can be recast in the  approximate form:
\begin{equation}
\frac{D H_{\rm R}}{Dt } \sim \mu \delta P_{\rm d}, 
\end{equation}
where $\delta$ is the lengthscale of variation of the magnetic field associated with the current density, i.e., $J \sim B/(\delta \mu)$, and $P_{\rm d}$ is the dissipated power. 
We assume that the  largest turbulent lengthscale is  $\delta \sim |\alpha|^{-1} = 0.12 R^{-1} = 6.7 \times 10^{9}$~m for $R = 8 \times 10^{8}$ m. Adopting  the maximum dissipated power suggested by the observations, i.e., $P_{\rm d} = 10^{21}$~W, we find:
\begin{equation}
\frac{D H_{\rm R}}{Dt} \sim 8.37 \times 10^{24}  \mbox{ T$^{2}$ m$^{4}$ s$^{-1}$}.
\label{hel_diss_rate}
\end{equation}

In the above derivation we have guessed the dissipated power from the available observations. For the mechanism to be at least plausible, we need to show that it is possible to sustain that level of dissipated power. This implies a computation of  the energy stored in the non-linear force-free field that is produced by the reconnection process, which is beyond the capability of our simple linear model. Nevertheless, we may estimate the energy available by means of a simplified argument that is by no means rigorous, but  has the advantage of using our  model for linear force-free configurations. Its results should be regarded only as an illustration of the plausibility of the proposed mechanism, deferring a more detailed and rigorous treatment to future studies. 

To compute the available energy, we assume that the reconnection between the coronal and planetary fields produces a dissipation of the helicity of the initial field configuration, i.e., that unperturbed by the planet. Again, we specify our considerations for the case of HD~179949, adopting the linear force-free model of \citet{Lanza08}. Therefore, we fix $\alpha=-0.12 R^{-1}$ and $b_{0}/c_{0}=-1.1$ in the initially unperturbed state, giving a total energy $E/E_{\rm p} = 1.0273$ and a relative helicity $|H_{\rm R}| = 55.3116 \; B_{0}^{2} R^{4}$ (cf. Eqs.~\ref{Benergy} and \ref{Bhelic}).  The boundary conditions at the stellar surface specify the values of $g^{\prime}(q_{0})$ and $\alpha $ through Eqs.~(\ref{BC_eqs}). The value of $\alpha$ depends on  the radial and azimuthal field components at the  surface and must be regarded as fixed. Conversely, the value of $b_{0}/c_{0}$ is pratically independent of the value of $g^{\prime} (q_{0})$ for $b_{0}/c_{0} < -1.0$. This is illustrated  in Fig.~\ref{gprime} and is discussed further in Sect.~\ref{intermittency}. In other words, the ratio $b_{0}/c_{0}$ is only marginally constrained by the boundary conditions and can be assumed to vary in order to decrease the total  helicity of the field. The minimum helicity is obtained for $b_{0}/c_{0} \rightarrow - \infty$, which gives $g^{\prime} (q_{0}) = -8.2133$ while for the initial field configuration with $b_{0}/c_{0}= -1.1$, $g^{\prime} (q_{0}) = -8.2021$, corresponding to a variation of the boundary condition by only 0.13 percent. This state of the field has $E/E_{\rm p} = 1.0180$ and $|H_{\rm R}| = 54.6421 \; B_{0}^{2} R^{4}$, i.e., an helicity  variation $\Delta H_{\rm R} = 0.67 B_{0}^{2} R^{4}$ with respect to the initial state. Assuming $B_{0}= 10$~G and $R = 8 \times 10^{8}$ m, the transition from the initial  state to this final state  releases an energy  $\Delta E = 0.0093 E_{\rm p} = 1.6 \times 10^{25}$ J. A crucial point is the timescale for the release of $\Delta E$ because it determines the power available to explain the observed phenomena. It depends on the time scale for the dissipation of the helicity and can be computed from the helicity dissipation rate as 
$\tau_{\rm hd} \sim \Delta H_{\rm R}/(dH_{\rm R}/dt) \sim 3.3 \times 10^{4}$~s, where we made use of Eq.~(\ref{hel_diss_rate}). 
Therefore, the available maximum power, estimated from $\Delta E/ \tau_{\rm hd}$, turns out to be $ \sim 4.9 \times 10^{20}$~W which is of the right order of magnitude to account for the chromospheric hot spot and the enhancement of X-ray flux. 

It is important to note that our argument provides us only with a lower limit for the energy that can be released because we approximate the process as a transition between two linear force-free fields. As a matter of fact, when the field is in a non-linear force-free state, its energy is  greater than the energy of the initially unperturbed linear force-free field, allowing the system to release more energy than we have estimated above. This can explain the modulation of the emitted flux with the orbital phase of the planet if only a portion of the stellar corona surrounding  the flux tube connecting the planet with the star is involved in the energy release process at any given time.   

The initial field configuration needs to be restored  on a time scale comparable with the orbital period of the planet to achieve a quasi-stationary situation in agreement with the observations.  
An order of magnitude estimate of the  helicity flux coming from the photospheric motions can be obtained  from Eq.~(11) of \citet{HeyvaertsPriest84} as:
\begin{equation}
\frac{d H_{\rm R}}{dt} \approx  4 \pi R^{2} B_{0}^{2} v_{\rm e} \alpha^{-1}, 
\label{hel_incoming}
\end{equation}
where $v_{\rm e}$ is the velocity of magnetic flux emergence at the photosphere, that we can take as a fraction, say 0.1, of the convective velocity, yielding $v_{\rm e} = 150 $ m s$^{-1}$. The timescale for restoring the helicity of the above magnetic configuration  is: 
\begin{equation}
\tau_{\rm hr} \equiv \frac{\Delta H_{\rm R}}{dH_{\rm R}/dt} \sim 3.2 \times 10^{4} \mbox{ s},
\end{equation}
where $dH_{\rm R}/dt$ comes now from Eq.~(\ref{hel_incoming}). It is 
comparable to the fastest helicity dissipation timescale applied above and is significantly shorter { (i.e., $\approx 10$ percent)} than the orbital period of  the planet. The magnetic energy flux associated with the helicity build up can be estimated as: $\sim 2 \pi R^{2} B_{0}^{2} v_{\rm e} / \mu \sim 4.8 \times 10^{20}$ W,
which can account for the observed energy release in the corona. 

%%%%%%%%%%%%%%%%%%%%%%%%%%%%%%%%%%%%%%%%%%%%%%%%%%%%%%%%%%%%%%%%%
%%%%%%%%%%%%%%%%%%%%%%%%%%%%%%%%%%%%%%%%%%%%%%%%%%%%%%%%%%%%%%%%%
{ It is interesting to note that the greater the magnetic field of a hot Jupiter, the more efficient the triggering of  helicity dissipation  in a stellar corona by the planet itself because the dissipated power scales as $B_{\rm pl}^{2/3}$ in Eqs.~(\ref{rec_power}) and (\ref{helic_diss_rr}). However, even if the planetary magnetic field is negligible, we still expect some dissipation by the currents induced in the planetary conductive interior by its motion through the stellar coronal field \citep{Laineetal08} or by the currents associated with the Alfven waves excited by the motion of the planet through the stellar wind \citep{Preusseetal05,Preusseetal06}. The corresponding helicity dissipation rate in Eq.~(\ref{helic_diss_rr}) is expected to be at least two orders of magnitude smaller than when the planet has a field $B_{\rm pl}$ of $5-10$~G, essentially because the current dissipation is confined within a much smaller volume. Therefore, the efficiency of the proposed mechanism is likely to be significantly reduced when the planetary field vanishes.  }

\subsection{The intermittent nature of SPMI}
\label{intermittency}

The fact  that chromospheric hot spots rotating sychronously with the planetary orbit have not always been observed in HD~179949 and $\upsilon$ And \citep{Shkolniketal08} can be interpreted by assuming that their  activity cycles are very short, i.e., of the order of $1-2$ years. If the large scale dipole field  cyclically reverses its direction,  we have alternate phases of strong and weak interactions with the magnetic field of the planet \citep[see ][]{Lanza08}. Such a hypothesis has  received some support from the observation of two consecutive field reversals in a couple of years  in $\tau$ Boo  \citep{Donatietal08,Faresetal09}. { Moreover, \citet{CranmerSaar07} and \citet{Shkolniketal09} build a statistical model for SPMI, based on the observations of the solar photospheric field along cycle 22 and its extrapolation to the corona by means of a potential field model, which predicts an intermittent interaction. The relative durations of the on and off phases depend on the phase of the solar cycle and are in general agreement with the few available observations reported in Sect.~\ref{observations}.  Nevertheless, another interpretation is possible which does not require a global reversal of the magnetic field or a complex geometry of the coronal field, but only a change of the photospheric boundary conditions that induces a change of the global topology of the coronal field. }

\begin{figure}[t]
\includegraphics[width=8cm,height=8cm]{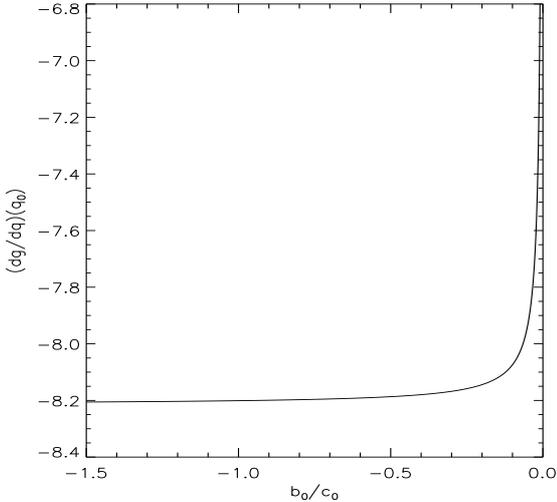} % {gp_surface.ps} 
\caption{The first derivative of the function $g$ at the surface of the star (i.e., $r=R$) vs. $b_{0}/c_{0}$ for $\alpha=0.12 R^{-1}$. }
\label{gprime}
\end{figure}

We shall explore the effect of a change of the boundary conditions on the force-free configuration adopted by \citet{Lanza08} to model SPMI in the case of HD~179949. It assumes $\alpha=-0.12 R^{-1}$ and 
$b_{0}/c_{0} = -1.1$. The mean of the meridional field component over the surface of the star 
$B_{\theta}^{(s)}$ 
in the r.h.s. of the second of Eqs.~(\ref{BC_eqs}) fixes the value of $g^{\prime}(q_{0})$ from which we can derive the model parameter $b_{0}/c_{0}$. A plot of $g^{\prime}(q_{0})$ vs. $b_{0}/c_{0}$ is given in Fig.~\ref{gprime} for $\alpha=-0.12 R^{-1}$ and shows that there is a large interval of 
$b_{0}/c_{0}$ where the variation of $g^{\prime}(q_{0})$ is very small. In other words, a modest variation of the meridional field component  produces a remarkable variation of $b_{0}/c_{0}$. On the other hand, to change significantly the parameter $\alpha$ we need a remarkable change of the axisymmetric azimuthal field component over the surface of the star that can be achieved only on timescales comparable with the stellar activity cycle. Therefore, the dependence on the boundary conditions suggests to explore the modification of the field topology for a variable  $b_{0}/c_{0}$ helding $\alpha$ fixed. 
%%%%%%%%%%%%%%%%%%%%%%%%%%%%%%%%%%%%%%%%%%%%%%%%%%%%%%%%%%%%%%%%%
%%%%%%%%%%%%%%%%%%%%%%%%%%%%%%%%%%%%%%%%%%%%%%%%%%%%%%%%%%%%%%%%%

\begin{figure}[t]
\includegraphics[width=8cm,height=12cm]{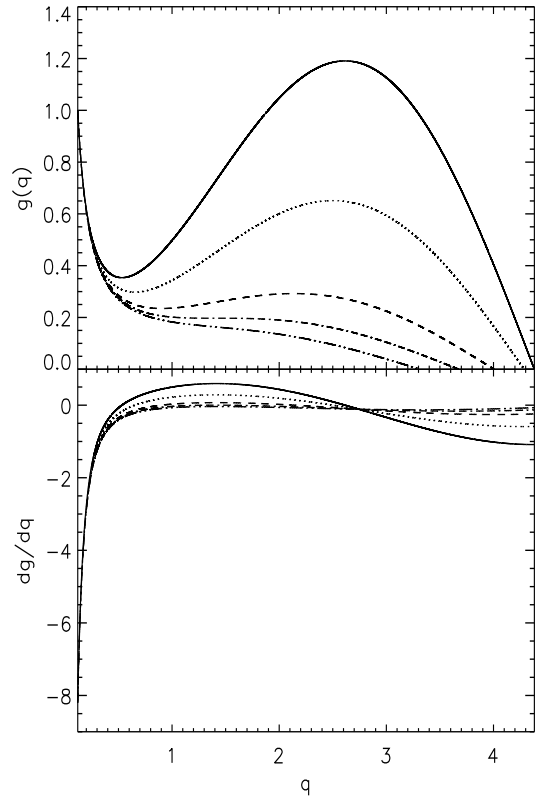} %{g_evol.ps} % {mag_hd179949_neu.ps}
\caption{{\it Upper panel}: The radial function $g$ vs. $q$ for $\alpha=-0.12 R^{-1}$ and different values of $b_{0}/c_{0}$ as indicated by the different linestyles, i.e., solid: $b_{0}/c_{0} = -0.107$; 
dotted: $-0.2$; dashed: $-0.5$; dash-dotted: $-0.9745$ (critical value for the formation of the azimuthal flux rope); dash-dot-dotted: $-2.0$. {\it Lower panel:} The first derivative $g^{\prime}$ vs. $q$ for $\alpha=0.12 R^{-1}$ and different values of $b_{0}/c_{0}$, according to the same linestyle coding adopted in the upper panel.} 
\label{gplots}
\end{figure}

\begin{figure}[t]
\includegraphics[width=8cm,height=8cm]{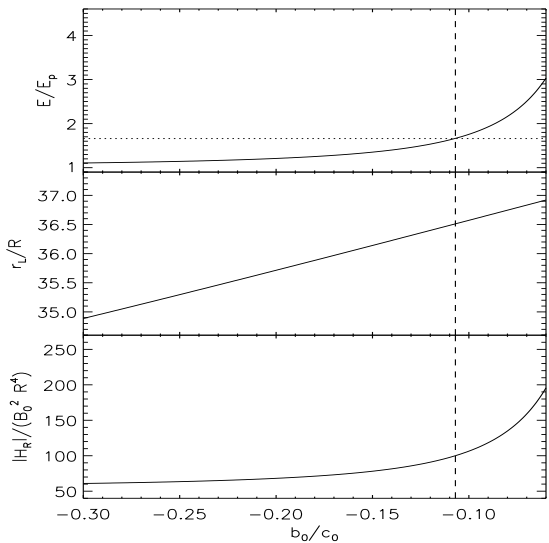} % {field_energy.ps} 
\caption{{\it Upper panel:} The magnetic field energy as a fraction of the corresponding potential field energy vs. $b_{0}/c_{0}$ for $\alpha=-0.12 R^{-1}$. The Aly energy limit is marked by the horizontal dotted line while the vertically dashed line marks the corresponding value of $b_{0}/c_{0}= -0.107$ and has been reported into all the other plots.
{\it Middle panel:} The outer radial limit of the field vs. $b_{0}/c_{0}$ for $\alpha=-0.12 R^{-1}$. {\it Lower panel:} the absolute value of the relative magnetic helicity vs. $b_{0}/c_{0}$ for $\alpha=-0.12 R^{-1}$.  }
\label{energ_plot}
\end{figure}

\begin{figure}[t]
\includegraphics[width=8cm,height=12cm]{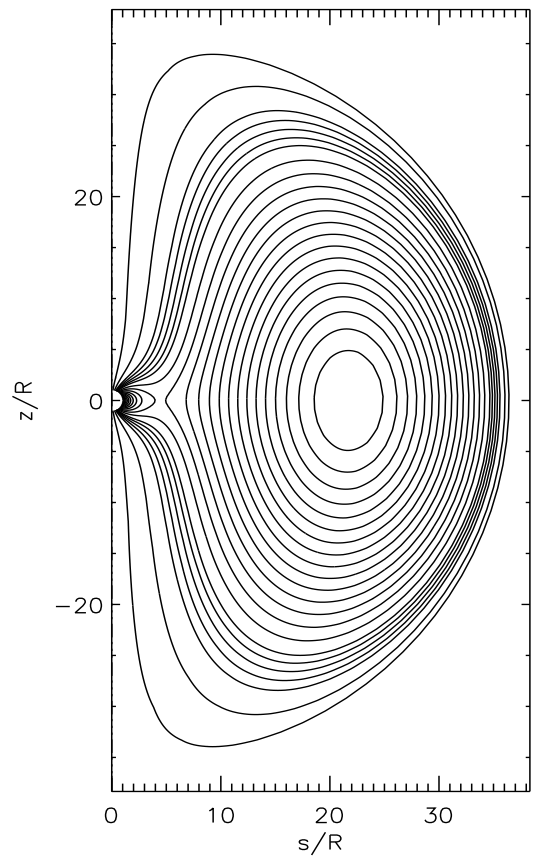} %{field_mer_sec1.ps} 
\caption{Meridional section of  the magnetic field lines for $\alpha=-0.12 R^{-1}$ and  $b_{0}/c_{0}=-0.107$ which corresponds to the Aly energy limit. The distance from the rotation axis is indicated by $s = r \sin \theta$, while the distance from the equatorial plane is $z = r \cos \theta$, with $R$ being the radius of the star. }
\label{mer_sec}
\end{figure}

The plots of the radial function $g(q)$ and its first derivative $g^{\prime}(q)$ vs. $q$ are shown in Fig.~\ref{gplots} for several values of $b_{0}/c_{0}$ and $\alpha=-0.12 R^{-1}$. For 
$(b_{0}/c_{0}) < - 0.9745 $, the function $g$ is monotonously decreasing and all the field lines have both ends anchored onto the photosphere. A sketch of the meridional section of the field lines is given in Fig.~1 of \citet{Lanza08} for $b_{0}/c_{0} = -1.1$. On the other hand, for $(b_{0}/c_{0}) \geq - 0.9745 $, the derivative $g^{\prime}$ vanishes at two points within the interval $q_{0} < q < q_{\rm L}$, and the field develops an azimuthal rope of flux centred around the maximum of $g(q)$ and whose inner radius coincides with the minimum of $g(q)$.  
%%%%%%%%%%%%%%%%%%%%%%%%%%%%%%%%%%%%%%%%%%%%%%%%%%%%%%%%%%%%%%%%%
%%%%%%%%%%%%%%%%%%%%%%%%%%%%%%%%%%%%%%%%%%%%%%%%%%%%%%%%%%%%%%%%%
The dependences of the magnetic field energy $E$, outer  radius $r_{\rm L}$, and absolute value of the relative helicity $|H_{\rm R}|$ on $b_{0}/c_{0}$ for $\alpha=-0.12 R^{-1}$ are illustrated in Fig.~\ref{energ_plot}. For $ -2.0 \leq b_{0}/c_{0} \leq -0.3$, the energy, outer radius, and absolute value of the relative helicity all increase very slowly with 
the  parameter, thus we restrict the plot to the interval showing the steepest variations. The  parameter corresponding to the Aly energy is $b_{0}/c_{0} = -0.107$ and it is marked by a vertically dashed line. 
%%%%%%%%%%%%%%%%%%%%%%%%%%%%%%%%%%%%%%%%%%%%%%%%%%%%%%%%%%%%%%%%%
%%%%%%%%%%%%%%%%%%%%%%%%%%%%%%%%%%%%%%%%%%%%%%%%%%%%%%%%%%%%%%%%%
A meridional section of the field lines when the field has the Aly energy is plotted in Fig.~\ref{mer_sec} and shows how the azimuthal flux rope has extended to occupy most of the available  volume, while the domain with field lines connected to the surface of the star has been squeezed below a radial distance $\sim 5 R$.
%%%%%%%%%%%%%%%%%%%%%%%%%%%%%%%%%%%%%%%%%%%%%%%%%%%%%%%%%%%%%%%%%
%%%%%%%%%%%%%%%%%%%%%%%%%%%%%%%%%%%%%%%%%%%%%%%%%%%%%%%%%%%%%%%%%

We have illustrated the dependences of $g^{\prime}$, field energy and helicity on the parameter $b_{0}/c_{0}$ for  the particular case of $\alpha=-0.12 R^{-1}$, but their qualitative behaviours are the same also for different values of $\alpha$. Specifically, we have explored numerically the dependence of $E/E_{\rm p}$ and $|H_{\rm R}|$ in the rectangle $(0.01 R^{-1}\leq | \alpha | \leq 0.6 R^{-1}) \times (-1.5 \leq b_{0}/c_{0} \leq -0.001)$ finding that they are always monotonously increasing functions of $b_{0}/c_{0}$ for any fixed value of $\alpha$. 

In Sect.~\ref{energy_budget2}, we have assumed that the helicity of the stellar corona is 
determined by a dynamical balance between the opposite contributions of 
the emerging magnetic fields and photospheric motions that build it up,  and the orbital motion of the planet that triggers a continuous dissipation of helicity and magnetic energy in the corona. Therefore, the photospheric boundary conditions that fix the value of the helicity in our model can be regarded as a result of those processes that rule the helicity balance of the stellar corona. 

To explain the transition between states with and without a chromospheric hot spots, we assume that the helicity of the coronal field and the corresponding boundary conditions can vary on a timescale shorter than the stellar activity cycle.  
Specifically, the flux of helicity into the corona and the meridional component of the surface field  are expected to vary as a result of the fluctuations characterizing turbulent hydromagnetic dynamos \citep[cf., e.g.,][]{BrandenSubra05}. As a consequence, $b_{0}/c_{0}$ varies, spanning a certain range of values. If such a range is large enough, the value of the parameter can sometimes cross the thresold $b_{0}/c_{0}= -0.9745$ for the transition to a flux rope topology, thus halting the SPMI mechanism described in Sect.~\ref{energy_budget2}. Conversely, when the parameter crosses the threshold  in the reverse direction, SPMI is resumed. { The relative durations of such on and off phases depend on the statistical distribution of the fluctuations of the meridional component of the surface field which is presently unknown. However, an on/off transition can take place on a time scale as short as $10^{5}-10^{6}$ s assuming the helicity fluxes and dissipation rates estimated in Sect.~\ref{energy_budget2}.} 

\begin{figure}[t]
\includegraphics[width=8cm,height=8cm]{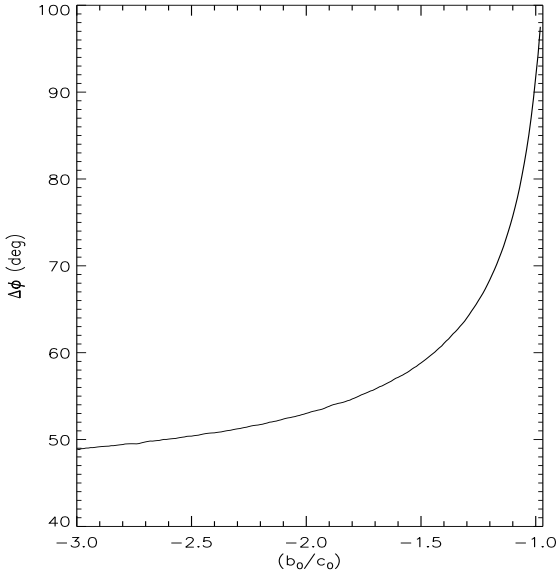} %{azimuthal_angle.ps} 
\caption{The phase lag between the planet and the synchronous chromospheric hot spot vs. $b_{0}/c_{0}$ for $\alpha=0.12 R^{-1}$. }
\label{delta_phi}
\end{figure}

The phase lag $\Delta \phi$ between the planet and the chromospheric hot spot depends on $b_{0}/c_{0}$ for a fixed $\alpha $ \citep[see ][ for the method to compute $\Delta \phi$]{Lanza08}. We plot such a dependence in Fig.~\ref{delta_phi}. 
 When the field has no flux rope, i.e., $ b_{0}/c_{0} < -0.97$, $\Delta \phi$ depends only slightly on $b_{0}/c_{0}$ varying only by $\pm 20^{\circ}$ around the mean observed value of $\sim 70^{\circ}$, { even for values of $b_{0}/c_{0}$ as small as $-3.0$, i.e., well beyond the range considered to model the observations of HD~179949}. The variation of $\Delta \phi$ becomes steep only when the field is very close to develop a flux rope. The inner radius of the flux rope reaches the distance of the planet ($r/R = 7.72$) when $b_{0}/c_{0}= -0.908$. We assume that the transition from, say, $b_{0}/c_{0}= -1.1$ to $-0.9$ occurs on a time scale so short that it has no observable consequences on $\Delta \phi$ given the limited duty cycles  of  present ground-based observations. 
%%%%%%%%%%%%%%%%%%%%%%%%%%%%%%%%%%%%%%%%%%%%%%%%%%%%%%%%%%%%%%%%%
%%%%%%%%%%%%%%%%%%%%%%%%%%%%%%%%%%%%%%%%%%%%%%%%%%%%%%%%%%%%%%%%%

\begin{figure}[t]
\includegraphics[width=8cm,height=8cm]{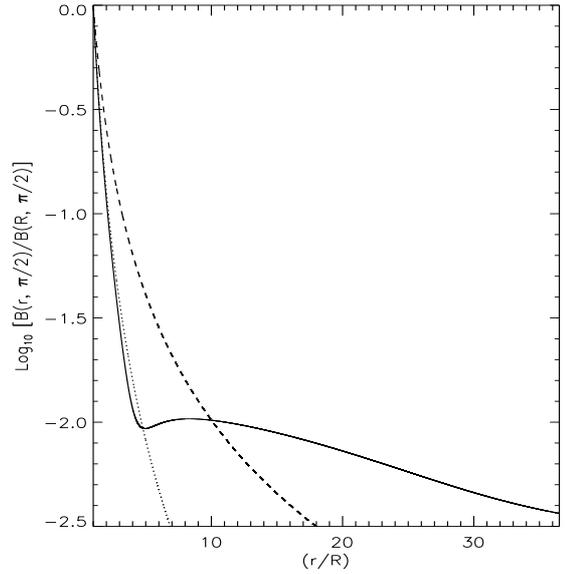} %{field_mod1.ps}
\caption{The modulus of the magnetic field on the equatorial plane of the star normalized to its value at the surface vs. the radial distance from the star for $b_{0}/c_{0}=-0.107$ and $\alpha=-0.12 R^{-1}$ (solid line). For comparison, we plot also the case of a potential field (dotted line) and of a field decreasing as $r^{-2}$ (dashed line). }
\label{magneticf2}
\end{figure}

By increasing $b_{0}/c_{0}$, the energy of the field grows up to the Aly limit where an instability opening the field lines may possibly be triggered.  The magnetic field intensity vs. the distance from the star is plotted in Fig.~\ref{magneticf2} for a model with $\alpha= -0.12 R^{-1}$ and the Aly energy. 
%%%%%%%%%%%%%%%%%%%%%%%%%%%%%%%%%%%%%%%%%%%%%%%%%%%%%%%%%%%%%%%%%
%%%%%%%%%%%%%%%%%%%%%%%%%%%%%%%%%%%%%%%%%%%%%%%%%%%%%%%%%%%%%%%%%
Following a rapid initial decrease, closely similar to that of a potential field, the field stays almost constant for $5 \leq (r/R) \leq 14$, and then decreases slowly far away from the star. This is due to the slow decrease of the field intensity with radial distance inside the flux rope. A first consequence is a remarkable increase of the  power dissipated at the boundary of the planetary magnetosphere, as given by Eq.~(\ref{rec_power}). In the case of   \object{HD~179949},  the field
intensity at $a =7.72R$ is  $\sim 3$ times greater than in the absence of the rope (cf. Fig.~\ref{magneticf1}), leading to an increase of $P_{\rm d}$ by a factor of $\sim 4$. Nevertheless, the released energy cannot reach the chromosphere and no hot spot is formed in the present case.
If a balance is eventually reached between the helicity flux from the photosphere and the dissipation in the corona, the flux rope configuration may become stationary. Its end may occur either by an increase of the energy above the Aly limit which may open up the field lines, or by a sudden decrease of the helicity flux which will change the topology of the field into one with all field lines connected to the stellar surface, thus resuming a chromospheric hot spot. 

The explanation suggested above for the intermittent behaviour of SPMI can be tested by measuring the direction of $B^{(s)}_{\theta}$, i.e., the meridional field component at the stellar surface, which is possible by means of spectropolarimetric techniques \citep[cf. ][]{Moutouetal07,Donatietal08} if the star rotates fast enough ($v \sin i \geq 12-15$ km s$^{-1}$).
If the on/off SPMI transition is not associated with a reversal of $B^{(s)}_{\theta}$, the present   explanation gains support. In principle, by measuring the  surface components of the field and the  angle $\Delta \phi$ with sufficient accuracy, it is possible to estimate $b_{0}/c_{0}$, thus  constraining  the model parameters during the SPMI on phases. Specifically, we can use the first and the third of Eqs.~(\ref{BC_eqs}) to find $B_{0}$ and $\alpha $, and then fix the range of $b_{0}/c_{0}$ that reproduces the observed $\Delta \phi$ (cf., e.g., Fig.~\ref{delta_phi}). The second of Eqs.~(\ref{BC_eqs}) can be used as an independent check of the accuracy of the linear force-free assumption because the resulting value of $g^{\prime}(q_{0})$ should be close to $-8$ if our model is indeed applicable. 

\subsection{Azimuthal flux rope and  planetary evaporation}

An interesting property of the flux rope topology is  the possibility of storing  matter in the stellar corona. The evaporation of hot Jupiters under the action of the ionizing radiation of their host stars originates a flow of cool plasma at a temperature of $\sim 10^{4}$ K that escapes from the planetary atmospheres into the stellar coronae \citep[cf., e.g., ][]{Ehrenreichetal08,Murray-Clayetal09}. The flux rope geometry keeps the evaporating plasma confined into a torus in the equatorial plane of the star, thus making its detection easier in the case of transiting hot Jupiters \citep{Vidal-Madjaretal03}. 

Considering a moderately active star and assuming a lifetime of $\sim 300$ days for the flux rope configuration according to the duration of the off phases of chromospheric interaction, 
the evaporated mass  is of the order of $5 \times 10^{14}$ kg \citep[cf., e.g., ][]{Murray-Clayetal09}.
Such a material may eventually condense in the form of  prominence-like structures around the minimum of the gravitational potential inside the flux rope, that is  in the equatorial plane of the star, possibly closer to the star than the hot Jupiter (cf. Fig.~\ref{mer_sec}). 
{ It is unlikely that  all the evaporated matter collects into a single condensation. The thermal instability of the plasma is expected to lead to many condensations with typical lengthscales of 
$10^{8}-10^{9}$~m, comparable in order of magnitude to those observed  in, e.g., solar prominences \citep{Field65}. Moreover, a high degree of inhomogeneity is expected inside each condensation, in analogy with the filamentary structure of solar prominences, because the  pressure scale height at a temperature of $\sim 10^{4}$~K is only of the order of $10^{5}-10^{6}$~m.} 

Assuming a mass of $1 M_{\odot}$ for the star and a radius $R=1 R_{\odot}$, the gravitational potential energy of the whole prominence material is  $\sim 2 \times 10^{25}$ J, for a mean radial distance of $7R$. It is only 1 percent of the magnetic energy of the coronal field, therefore it is not expected to perturb appreciably the field geometry. 

 On the other hand, in the case of a very active star, i.e., {having an X-ray luminosity $2-3$ orders of magnitude greater than the Sun \citep[cf., e.g., ][]{Penzetal08}}, the planet evaporation rate may increase by a factor of $\sim 100$, leading to a potential energy of the whole prominence material of the order of $10^{27}-10^{28}$~J. This can significantly perturb the field, whose configuration can no longer be assumed force-free. A non-force-free model may be applicable, such as that of \citet{Neukirch95},  discussed by \citet{Lanza08}. It gives rise to the same field topologies of the present force-free model because it is obtained from a linear force-free model by means of a linear transformation of the independent variable $q$ of the radial function $g(q)$ \citep[cf. ][]{Lanza08}.
A large amount of prominence material may help to stabilize the flux rope configuration, because lifting up the prominences to open or change the geometry of the magnetic field lines  requires a comparable additional amount of energy. This may imply that the flux rope topology is the most stable in the case of very active stars, leading to a lower probability of observing a chromospheric hot spot synchronized with the planet. 

Prominence-like structures have indeed been detected around some young and highly active dwarf stars, through the absorption transients migrating across their H$\alpha$, Ca~II H \& K and Mg~II h \& k emission line profiles \citep[e.g., ][]{CameronRobinson89}. Typical masses are in the range $(2-6) \times 10^{14}$ kg  for the { young ($\approx 50$ Myr)}, single K0 dwarf AB~Dor   \citep{Cameronetal90}. We expect a remarkable increase of the number of such phenomena in highly active stars hosting evaporating hot Jupiters. Assuming a lifetime of $\sim 300$ days for the flux rope configuration, a total evaporated mass of $2 \times 10^{16}$ kg can lead to the formation of 
$\sim 100$ prominence-like structures akin those observed in AB~Dor.

The most favourable geometrical conditions for their detection are found in transiting systems where the prominences are expected to transit across the disc of the host star. The radial distance of a condensation inside a flux rope is of several stellar radii, in contrast to a minimum distance of a few stellar radii observed in the case of the prominences formed by evaporation from the chromosphere of the star \citep{CameronRobinson89}. This implies that the transit times of the absorption features across the line profiles  are shorter by a factor of $2-5$ in the case of  flux rope condensations. 
{ On the other hand, the expected relative decrease of the intensity along the H$\alpha$ line profile is between 5 and 20 percent, with typical equivalent widths of the absorption features between 50 and 500 m\AA, as reported by \citet{CameronRobinson89} for AB~Dor,  whose coronal parameters can be considered fairly typical of those of very active stars.} 

Finally, note that a prominence does not form at the distance of the planet when the field topology does not contain a flux rope because in that case all the evaporated matter will fall onto the star, except when the star rotates so fast that the centrifugal force  reverses the effective gravity at the top of the loops interconnecting the planet with the star. 

\subsection{Consequences for X-ray emissions}

{  In our model, all the magnetic field lines are closed, so the corona consists of closed loops. However, if we adopt a more realistic model, such as the non-linear force-free model of \citet{Flyeretal04}, it is possible to have open field lines which may account for the configuration observed in solar coronal holes. Also in those non-linear  models we can have a coronal field with a large azimuthal flux rope which  affects the topology of the coronal field lines close to the star. From a qualitative point of view, we can refer to  Fig.~\ref{mer_sec} showing that  all the magnetic structures at low and intermediate latitudes must have a closed configuration with a top height lower than $\sim 4-5 \; R$ because they must lie below the flux rope. This implies that open field  configurations, akin solar coronal holes, are not allowed at low latitudes when a sizeable flux rope has developed in the outer corona. Since closed magnetic configurations are characterized by X-ray fluxes up to $\sim 10-100$ times greater than coronal holes, an azimuthal flux rope configuration is expected to be associated with a greater X-ray luminosity of the star than in the case when all magnetic field lines are connected to the photosphere. Of course, this is independent of the presence of a planet. 

When a star is accompanied by a close-in planet, it may increase the dissipation of magnetic helicity and energy, as conjectured in Sect.~\ref{energy_budget2}. According to \citet{Kashyapetal08}, stars with a distant planet have an average X-ray luminosity $L_{\rm X} \sim 7 \times 10^{20}$ W, while stars with a hot Jupiter have $L_{\rm X} \sim 5.5 \times 10^{21}$ W.
Considering the model introduced in Sect.~\ref{energy_budget2}, the dissipated power is proportional to $B_{0}^{2}$, where $B_{0}$ is the photospheric magnetic field (see Sect.~\ref{model}). This holds true both for a field topology with all field lines connected to the photosphere and with an azimuthal flux rope. We find that for both field topologies an average photospheric field $B_{0} \sim 30$~G is sufficient to account for the enhancement of X-ray luminosity in stars with hot Jupiters.  Since our model underestimates the free energy available in real non-linear force-free fields, such a value of $B_{0}$ should be regarded as an upper limit, thus our estimate agrees well with the available observations \citep[e.g., ][]{Moutouetal07}.}

\subsection{Consequences for radio emission}

The frequency of  radio emission from exoplanets depends on { the planetary} magnetic fields
\citep{Zarka07}.
According to the scaling laws adopted by \citet{Griessmeieretal04}, the magnetic moments of hot Jupiters are expected to be  one order of magnitude smaller than that of Jupiter owing to  tidal synchronization between their rotation and orbital motion. Adopting a surface field of 10 percent of Jupiter, i.e., 1.4 G, the cyclotron maser emission should peak at $\sim 4$ MHz, which is impossible to detect  because it  falls below the plasma frequency cutoff \citep{JardineCameron08}. The situation is much more favourable if the magnetic fields of hot Jupiters are indeed as strong as predicted by the models of \citet{Christensenetal09} in which the planetary dynamo is powered by the internal convective motions and the field intensity is pratically independent of the rotation rate of the planet. Assuming a field intensity at the surface of $25$ G, the  cyclotron maser emission peaks at frequencies around 70 MHz. 

In addition to the  emission from the poles of the planet, we expect radio emission from the stellar corona. When the field lines are connected to the stellar surface, electrons accelerated at the reconnection sites  travel down to the star producing emissions up to the GHz range from localized regions above photospheric spots with  fields of $10^{2}-10^{3}$ G. However, this is not possible when the field has a topology with an azimuthal flux rope. In this case the coronal field intensity is of the order of $10^{-2}$ G and the emission peaks at very low frequencies, comparable to or below the plasma frequency, i.e., it is self-absorbed before escaping from the emitting region. In this case, we expect detectable radio emission only from the poles of the planet, i.e., from a highly localized region. The beaming of cyclotron-maser emission may additionally decrease the power toward the observer thus explaining the lack of detection at frequencies of $50-100$ MHz. 

In conclusion, we expect that the best chances of detecting radio emission from exoplanets with the current instrumentation can be achieved when observing systems with a chromospheric hot spot synchronized with the planet. In this case, a fraction of the electrons accelerated at the reconnection sites inside the long loop connecting the star with the planet may be driven to the polar regions of the planet and to the stellar surface  increasing the irradiated power. In this case, the emission is expected to be strongly modulated with the  orbital period of the planet.

\section{Discussion}
\label{discussion}

We have presented a model for the coronal magnetic field of late-type stars that allows us to investigate the magnetic interaction between a star and a close-in exoplanet. The same model was used  by \citet{Lanza08} to explain the observed phase lags between  chromospheric hot spots attributed to SPMI and the planets. Now, we have discussed the energy budget of  SPMI and its intermittency on the base of that model. We suggest that the coronal field evolution in a star hosting a hot Jupiter is ruled by a dynamical balance between the helicity  coming up into the corona from the photosphere and that dissipated by the reconnection events triggered by the orbital motion of the planet. At any given time, most of the energy is dissipated in the loop connecting the planet with the stellar surface and in its neighbour magnetic structures rather than at the boundary of the planetary magnetosphere. 

This scenario is quite different from that characteristic of stars with distant planets, like our Sun. In that case the accumulation of magnetic helicity and energy in the corona leads to an instability of closed field structures that erupt as coronal mass ejections (CMEs) eliminating the excess of helicity \citep{ZhangLow05,Zhangetal06}. In the present model, in addition to the CME mechanism, the interaction between the coronal field and the planetary magnetosphere  takes part in the helicity dissipation process. In principle, one expects that a large flare that extends over most of the  stellar corona may sometimes occur  thanks to the capability of the planet to trigger a large-scale helicity dissipation process. If most of the coronal helicity is  dissipated during such an event, the maximum available energy can approach the difference between the Aly limit and the energy of the potential field with the same radial 
component at the photosphere, i.e., $\Delta E_{\rm max}= 0.66 E_{\rm p}$. For the case of HD~179949 with { an assumed} photospheric field of $B_{0}=10$ G, we have $\Delta E_{\rm max} = 1.1 \times 10^{27}$ J. Assuming a  photospheric mean field $B_{0}=40$~G and $R=0.75$ R$_{\odot}$, as suggested by the observations of  the K dwarf \object{HD~189733} by \citet{Moutouetal07}, we get $\Delta E_{\rm max} = 5.2 \times 10^{27}$ J. Such large energies may  explain the superflares observed in some dwarf stars, giving support to  a conjecture by  \citet{RubenScha00}.  

Processes like  CMEs, i.e., capable of reducing the helicity of the stellar field,  are required  for the operation of a stellar hydromagnetic dynamo \citep{BlackmanBranden03,BrandenSubra05}. Therefore, a hot Jupiter may help the star to get rid of the helicity generated by its dynamo in the convection zone, increasing dynamo efficiency and the overall level of magnetic activity. This may ultimately explain the greater X-ray luminosity of stars with hot Jupiters. { Moreover, shorter activity cycles may be expected \citep[cf., e.g., \S~11.2.1 of][]{BrandenSubra05}, which might account for the 2-year magnetic cycle suggested by the latest observations of $\tau$~Boo \citep{Faresetal09}. }

 The modulation of the helicity loss with the orbital period of the planet might account for the photospheric cool spots that appear to rotate synchronously with the planet, as conjectured by \citet{Lanza08}. These spots are related to the emergence of magnetic flux from the convection zone that may contribute to the formation of hot spots in the chromosphere and in the corona by reconnecting with  pre-existing fields, thus contributing to the energy budget of SPMI. 

Our model predicts a field intensity that decreases slower than that of a potential field far away from the star. In the case of the flux rope topology,  the field decreases even slower than $r^{-2}$ in the outer part of the closed corona. This may enhance the dynamical coupling between the star and the planet, as suggested by some preliminary computations of the angular momentum exchange between CoRoT-4a and its hot Jupiter having an orbital semimajor axis $ a \sim 17.4 R$, { where $R$ is the stellar radius}. They indicate that the rotation of the outer convection zone of the star may have been synchronized with the orbital motion of the planet if the surface field $B_{0} \geq 10$~G 
and the age of the system is $ \geq 0.5$ Gyr. These results may  explain the tight synchronization  observed in the CoRoT-4 system which is impossible to understand with tidal models given the large distance of the hot Jupiter and its low mass 
\citep[$ \sim 0.7$ M$_{\rm J}$, ][]{Lanzaetal09b}. 

A limitation of the present approach is the use of a linear force-free model for the stellar coronal field. This  mainly affects our energy estimates, while the main topological features of our linear model are shared  by non-linear models \citep[cf., e.g., ][]{Flyeretal04,Zhangetal06}. The latter can be useful to treat  the 
contribution of the outer coronal fields because they can provide us with field configurations that extend to the infinity with a finite magnetic energy.  
However, given the much greater mathematical complexity of non-linear force-free models,  the present treatment is preferable for a first description of the most relevant physical effects. 

Non-linear models can be useful also to investigate  the stability of coronal magnetohydrostatic configurations. In the framework of the adopted model,  stability is warranted by   Woltjer theorem because a linear force-free configuration is the minimum-energy state for  given total helicity and boundary conditions \citep{Berger85}. However, a real coronal field that is in a non-linear force-free state may become unstable before reaching the Aly energy limit by, e.g., kink modes, when it  develops an azimuthal flux rope.

\section{Conclusions}
\label{conclusions}

We have further investigated the model proposed by \citet{Lanza08} to interpret the observations of star-planet magnetic interaction. A linear force-free model of the stellar coronal field has been applied to address the energy budget of the interaction and to understand its intermittency.
We propose that the magnetic helicity budget plays a fundamental role in the interaction. 
An hot Jupiter contributes to this budget by increasing the helicity dissipation that triggers an  additional magnetic energy release in the stellar corona.

The transition between phases with and without a chromospheric hot spot rotating synchronously with the planet is interpreted as a consequence of a topological change of the coronal field  induced by an accumulation of helicity. { Its timescale depends on the mechanisms ruling the helicity fluctuations in the stellar hydromagnetic dynamo and the helicity budget of the stellar corona which are presently poorly known. However, a timescale as short as $10^{5}-10^{6}$ s could in principle be possible (cf. Sect.~\ref{intermittency}).  } 

The model can be tested in the case of sufficiently rapidly rotating stars, such as \object{$\tau$ Boo} or \object{HD~189733}, by combining spectropolarimetric  measurements of the stellar photospheric fields with Ca~II~K line observations to determine the phase lag between 
{ a planet-induced hot spot and the planet. Such simultaneous measurements can in principle constrain the topology of the field by allowing us to estimate the parameters of the force-free field model as detailed at the end of  Sect.~\ref{intermittency}}. 

The present model also bears interesting consequences for the coronal emissions of  stars hosting hot Jupiters. It may explain why stars with a close-in giant planet have, on the average, a higher X-ray luminosity than stars with a distant planet. Moreover, it suggests that the best chances to detect radio emission from hot Jupiters or their host stars are found in systems showing a chromospheric hot spot rotating synchronously with the planet. 

We have also investigated the consequences of the different field topologies for the confinement and the storage of the matter evaporated from a planetary atmosphere under the action of the  radiation from the host star. When the field has an azimuthal rope of flux encircling the star, the evaporated matter can neither escape nor fall onto the star and is expected to condense in the outer corona forming several prominence-like structures. It can in principle be detected in the case of rapidly rotating and highly active stars through the observations of transient absorption features moving across the profile of their chromospheric emission lines. 

{ Simultaneous optical, X-ray and radio observations can prove the association between a chromospheric hot spot induced by a hot Jupiter and the X-ray and radio emission enhancements  expected on the basis of our model because the energy is mainly released in the corona of the star and then conveyed  along magnetic field lines to heat the lower chromospheric layers. Most of the coronal energy should be released within one stellar radius, where the magnetic field is stronger. Therefore, no large phase lags are expected between the chromospheric enhancement and the X-ray and radio enhancements varying in phase with the orbital motion of the planet. On the other hand, when there is no signature of SMPI in the chromosphere, we expect that also the modulation of the X-ray flux with the orbital motion of the planet is significantly reduced. No detectable radio emission is expected in this case, except when the planet has a polar field of at least $20-30$ G. Finally,  we expect to observe the signatures of several prominence-like condensations in  the coronae of rapidly rotating ($v \sin i \geq 40-50$ km s$^{-1}$), highly active stars ($L_{\rm X} \sim 10^{22}-10^{23}$ W) hosting transiting hot Jupiters, if our assumption that coronal flux ropes have a typical lifetime of $200-300$ days is true.  }

\begin{acknowledgements}
The author wishes to thank an anonymous Referee for a careful reading of the manuscript and valuable comments. AFL is grateful to Drs. P.~Barge, S.~Dieters, C.~Moutou and I.~Pagano for drawing his attention to the interesting problem of star-planet magnetic interaction and for interesting discussions. This work has been partially supported by  the Italian Space Agency (ASI) under contract  ASI/INAF I/015/07/0,
work package 3170. Active star research and exoplanetary studies at INAF-Catania Astrophysical Observatory and the Department of Physics
and Astronomy of Catania University is funded by MIUR ({\it Ministero dell'Istruzione, Universit\`a e Ricerca}), and by {\it Regione Siciliana}, whose financial support is gratefully
acknowledged. 
This research has made use of the ADS-CDS databases, operated at the CDS, Strasbourg, France.
\end{acknowledgements}

\appendix

\end{document}